\documentclass[3p]{elsarticle} 

\usepackage{hyperref}
\hypersetup{colorlinks=true, urlcolor=blue}
\usepackage{subfigure}
\usepackage{epsfig}
\usepackage{newlfont}
\usepackage{amssymb}
\usepackage{amsfonts}
\usepackage{amsmath}
\usepackage{bm}

\def\lsim{\mathrel{\rlap{\lower4pt\hbox{\hskip1pt$\sim$}}
    \raise1pt\hbox{$<$}}}                
\def\gsim{\mathrel{\rlap{\lower4pt\hbox{\hskip1pt$\sim$}}
    \raise1pt\hbox{$>$}}}                

\DeclareMathOperator{\Tr}{Tr}

\journal{Physics Letters A}

\begin{document}

\title{Cumulative quantum work-deficit versus entanglement in the dynamics of an infinite spin chain}

\author[sps]{Himadri Shekhar Dhar}

\author[sps,snu]{Rupamanjari Ghosh}

\author[hri]{Aditi Sen(De)}

\author[hri]{Ujjwal Sen \corref{cor1}}
\ead{ujjwal@hri.res.in}

\cortext[cor1]{Corresponding author}

\address[sps]{School of Physical Sciences, Jawaharlal Nehru University, New Delhi 110067, India}

\address[snu]{School of Natural Sciences, Shiv Nadar University, Gautam Budh Nagar, UP 203207, India}

\address[hri]{Harish-Chandra Research Institute, Chhatnag Road, Jhunsi, Allahabad 211019, India}

\date{\today}

\begin{abstract}
We find that the dynamical phase transition (DPT) in nearest-neighbor bipartite entanglement of time-evolved states of the anisotropic infinite quantum XY spin chain, in a transverse time-dependent magnetic field, can be quantitatively characterized by the dynamics of an information-theoretic quantum correlation measure, namely, quantum work-deficit (QWD). We show that only those nonequilibrium states exhibit entanglement resurrection after death, on changing the field parameter during the DPT, for which the cumulative bipartite QWD is above a threshold. The results point to an interesting inter-relation between two quantum correlation measures that are conceptualized from different perspectives.
\end{abstract}



\maketitle

\section{Introduction}

Correlations 
form important indicators of the physics of a system involving a large number of interacting particles.
The behavior of quantum correlations \cite{4, KM} in macroscopic phases of such many-body systems
is potentially important
in understanding nonclassical properties such as critical phenomena, and quantum fluctuations \cite{1, 5}.
In particular, entanglement \cite{4} has been widely used in many-body physics to study near-critical behavior, phase transitions, and the general evolution in spin systems \cite{5}.

In recent years, the concept of nonclassical correlations in quantum systems has been taken beyond quantum entanglement, due to the existence of interesting few-body quantum phenomena
where entanglement is absent \cite{interest}.
This has led to the formulation
of information-theoretic quantifiers of quantum correlation, independent of the entanglement-separability criteria, 
based on the thermodynamics of local and global measurement strategies \cite{QD, QWD}. 
These measures have been used to study various aspects of quantum information, and in particular,
have been applied to investigate many-body phenomena such as
quantum phase transitions \cite{discord-spin}, correlation dynamics in many-body systems \cite{work2} (for a review, see \cite{KM})
and in open quantum systems \cite{patraghat}.
An important example of such an information-theoretic measure of nonclassical correlations is the quantum work deficit \cite {QWD}.

In this article, we study the behavior of quantum work deficit (QWD) \cite{QWD} for an infinite anisotropic quantum XY spin chain in a transverse time-dependent magnetic field. QWD is the difference in negentropy (``work'') that can be extracted by using global and local heat engines \cite{QWD,8}. The concept of QWD is based on the fact that information can be treated as a thermodynamic resource \cite{10} and it is hence defined as the difference between the amount of pure states that can be extracted under global operations and that under certain local operations. The difference accounts for the missing information (``resource") that can be attributed to the presence of
quantum correlations
in the system, independent of entanglement.

The nearest-neighbor entanglement of the nonequilibrium time-evolved state of an anisotropic infinite XY spin chain in a transverse time-dependent field is known to exhibit a 
dynamical phase transition (DPT) \cite{11} for a fixed, short time. Specifically, entanglement dies for a small initial magnetic field while the state becomes entangled with the increase of magnetic field.
We find that the nature of the DPT of bipartite entanglement can be characterized by studying the dynamics of QWD. In particular, we show that entanglement death and possible resurrection with changing initial magnetic 
field as observed during the DPT, can be inferred by the ``cumulative'' QWD in the system during the evolution. Only those nonequilibrium states exhibit revival after death for which the cumulative QWD is above a threshold value.
The results are independent of the anisotropy in the quantum XY model.

The quantitative relation exhibited between QWD and entanglement is potentially an interesting observation. 
Although 
for general mixed quantum states, there may not exist any direct relation between information-theoretic and entanglement measures (cf. \cite{comp}), in these specific class of systems, we establish a quantitative relation between them. The results may be of
interest in investigating the interplay of such quantum correlation measures in the dynamics of generic quantum many-body systems.   
%


\section{The XY spin chain}
\label{s1}

The Hamiltonian for the quantum spin chain that we consider is given by
$H(t)= H_{int} - h(t)H_{mag}$, where the external magnetic field has the form $H_{mag}= \sum_i S^z_i$.
The interaction term is given by
$
H_{int}=J\sum_i (\cal{A}S^x_iS^x_{i+1} + \cal{B}S^y_iS^y_{i+1}) ,
$
where $J$ measures the interaction strength, and $S^{j} = \frac{1}{2}\sigma^j$ ($\textit{j=x,y,z}$) are one-half of the Pauli spin matrices at the corresponding site. The spin coupling constants can be defined in terms of the anisotropy parameter $\gamma$ as $\cal{A}= 1+ \gamma$ and $\cal{B}= 1-\gamma$, where $\gamma \neq$ 0. The anisotropy parameter is chosen to ensure $[H_{int},H_{mag}] \neq 0$, and hence the external field evokes a non-trivial response in the system dynamics.
The external transverse field is applied in the form of a finite initial perturbation
at $t=0$: $h(0)=a > 0$, for $t=0$. The external field vanishes at $t>$ 0: $h(t) =0$, for $t>0$. Hence, the dynamics of the system at any time $t$ depends on the initial field $a$.
Note that $J$ and $a$ have the units of energy, while $\gamma$ is dimensionless.

To investigate the nearest neighbor (NN) correlation properties of the anisotropic XY spin chain, we require the two-site reduced density matrix of the nonequilibrium time-evolved state. 
If we consider the initial state of the system as a canonical equilibrium state at temperature $T$, the nonequilibrium two-site reduced density matrix, \(\rho_{12}^\beta\), at time $t$ is given by \cite{15}
\begin{eqnarray}
\rho_{\textit{12}}^{\beta}(t)&=& \frac{1}{4} \left[I \otimes I + M^z (t)( \sigma^z \otimes I + I \otimes \sigma^z) \right.\nonumber\\
&+& l^{xy}(t) (\sigma^x \otimes \sigma^y + \sigma^y\otimes \sigma^x) \nonumber\\
&+& \left.\sum_{j=x,y,z} l^{jj}(t) \sigma^j \otimes \sigma^j]\right.,
\label{twosite}
\end{eqnarray}
where $l^{ij}(t)~ (i,j=x,y,z)$ are the NN classical correlation functions and $M^z$ is the transverse magnetization. Upon diagonalizing the Hamiltonian, using the Jordan-Wigner and Fourier transformations, 
the NN classical correlation functions and the magnetization in Eq.\,(\ref{twosite})
are given as follows \cite{15}:
\begin{eqnarray}
l^{xy}(t)&=&l^{yx}(t)=s(t), \nonumber\\
l^{xx}(t)&=& g(-1,t),~ l^{yy}(t)=g(1,t),
\end{eqnarray}
where $g(i',t)$ (for $i'= \pm 1$) and $s(t)$, for initial temperature $T$=0, are given by
\begin{eqnarray}
g(i',t)&=&\frac{\gamma}{\pi}\int^\pi_0d\phi \frac{\sin(i'\phi)\sin \phi} {\Lambda(\tilde{a})\Lambda^2(0)}
\times\left\{\gamma^2 \sin^2\phi + (\cos\phi\right.\nonumber\\
&-& \tilde{a})\left.\cos\phi + \tilde{a}\cos\phi\cos[2\Lambda(0)\tilde{t}]\right\}-\frac{1}{\pi}\int^\pi_0d\phi\nonumber\\
&\times& \frac {\cos\phi} {\Lambda(\tilde{a}) \Lambda^2(0)}
  \left( \left\{\gamma^2\sin^2\phi + \left(\cos\phi-\tilde{a} \right) \cos\phi\right\}\right.\nonumber\\\nonumber\\
&\times& \left.  \cos\phi -\tilde{a}\gamma^2\sin^2\phi\cos[2\Lambda(0)\tilde{t}]\right),
\end{eqnarray}
\begin{equation}
s(t) = - \frac{\gamma \tilde{a}}{\pi} \int^\pi_0d\phi \sin^2 \phi \frac{\sin[2\tilde{t}\Lambda(0)]}{\Lambda(\tilde{a}) \Lambda(0)}.
\end{equation}
The magnetization is given by
\begin{eqnarray}
M^z(t) &=& \frac{1}{\pi} \int_0^\pi d\phi \frac{1}{\Lambda(\tilde{a}) \Lambda^2(0)}
     \times \{\cos[2 \Lambda(0)\tilde{t}] \gamma^2 \tilde{a}\sin^2\phi]\nonumber\\
  &-&\cos\phi [(\cos\phi - \tilde{a})\cos\phi + \gamma^2 \sin^2 \phi]\}.
\end{eqnarray}
Here, \(\Lambda(x)= \left\{\gamma^2\sin^2\phi~+~[x-\cos\phi]^2\right\}^{\frac{1}{2}}\), and 
\(\tilde{a} = a/J, \quad \tilde{t} = Jt/\hbar\).
In our analysis, the dimensionless variables $\tilde{a}$ and $\tilde{t}$ are used as the initial field and time parameters, respectively.

\begin{figure*}[t]
\begin{center}
\subfigure[]{\includegraphics[width=0.3\linewidth,angle=-90,clip=]{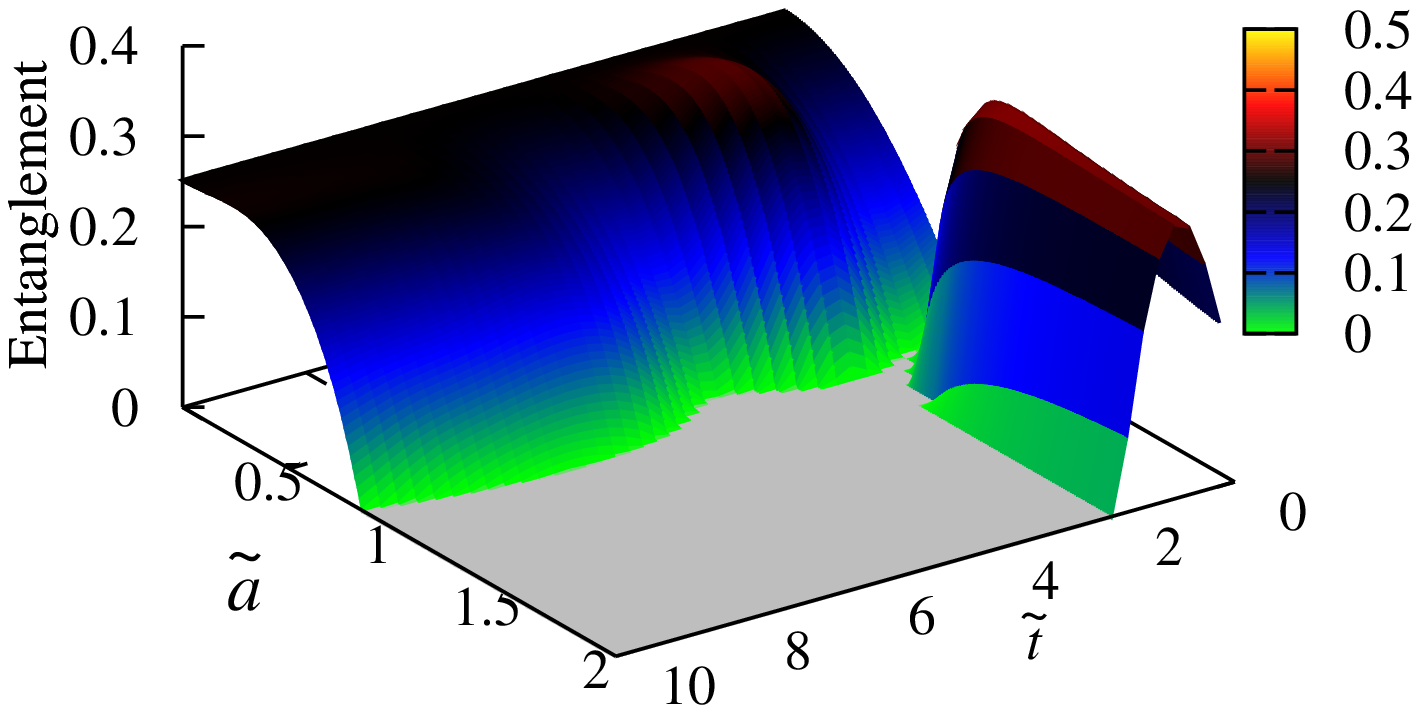}}\hspace{0.3cm}
\subfigure[]{\includegraphics[width=0.3\linewidth,angle=-90,clip=]{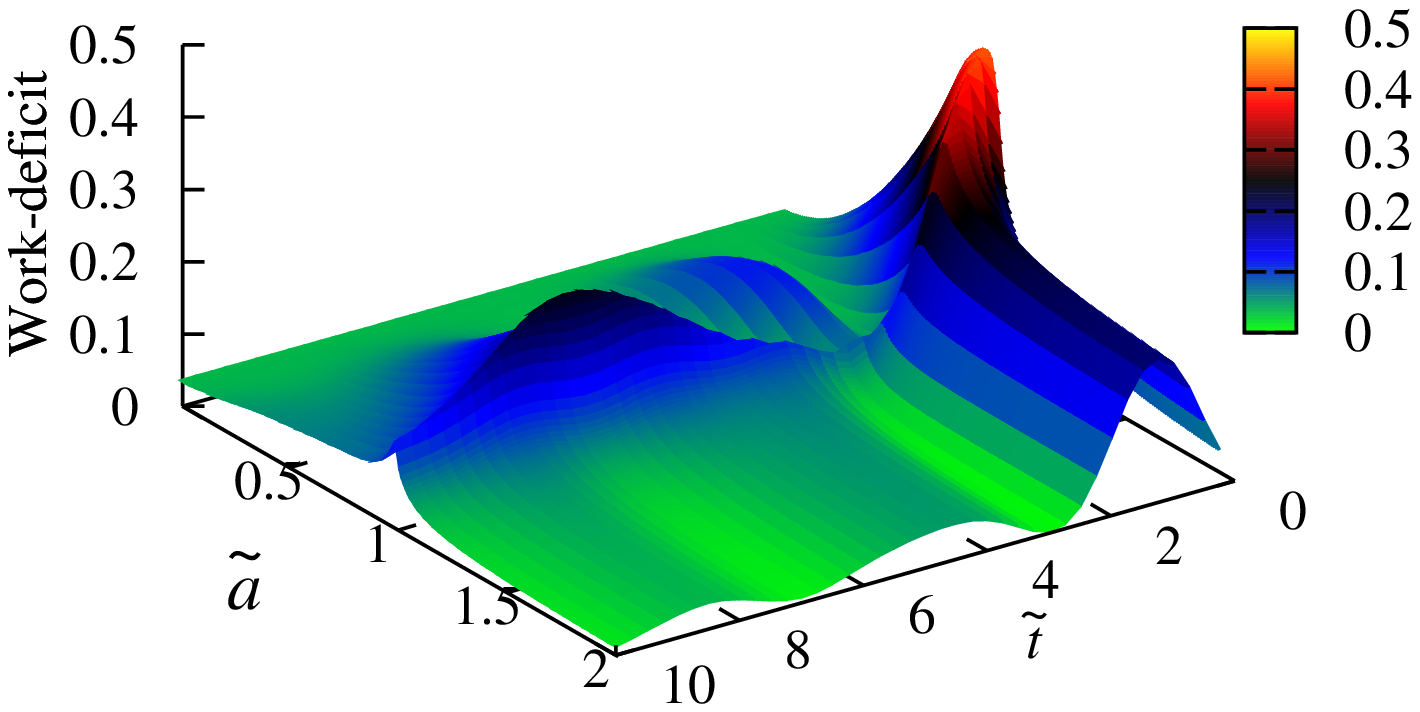}}\hspace{0.3cm}
\subfigure[]{\includegraphics[width=0.3\linewidth,angle=-90,clip=]{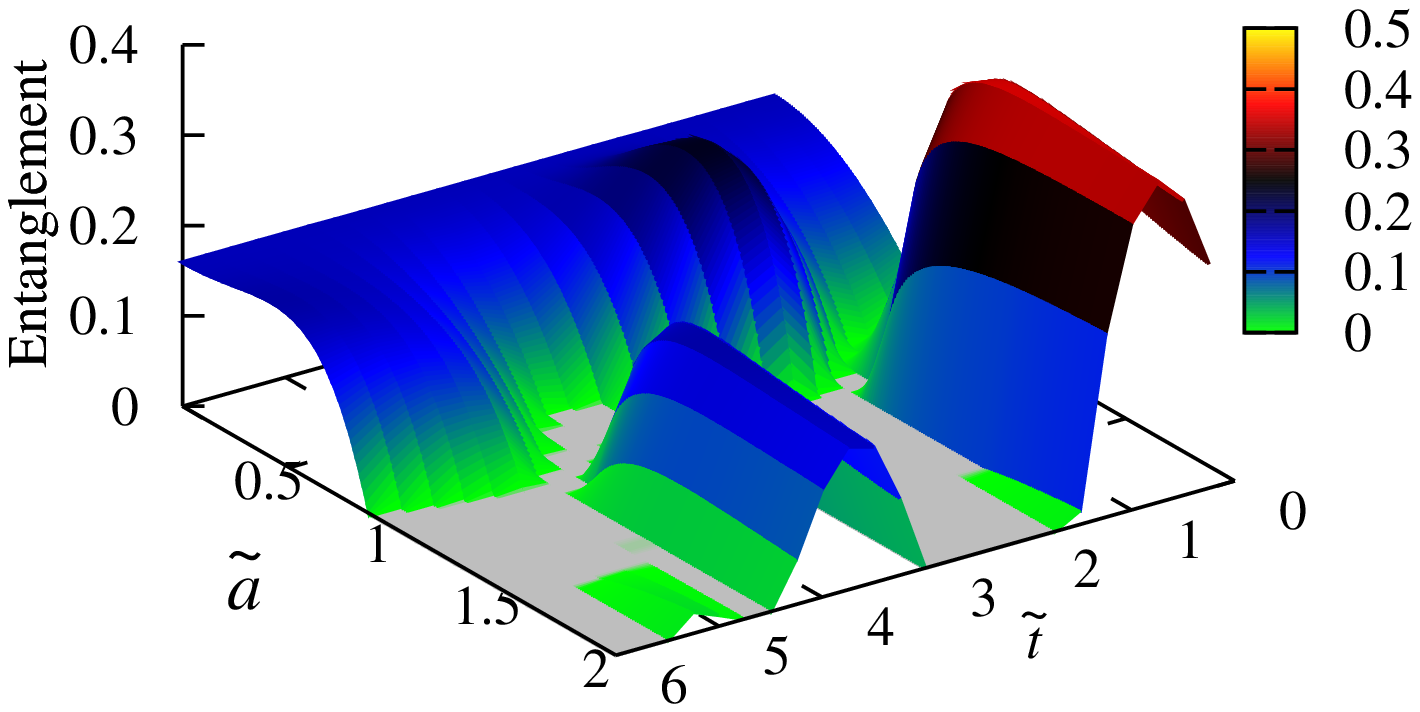}}\hspace{0.3cm}
\subfigure[]{\includegraphics[width=0.3\linewidth,angle=-90,clip=]{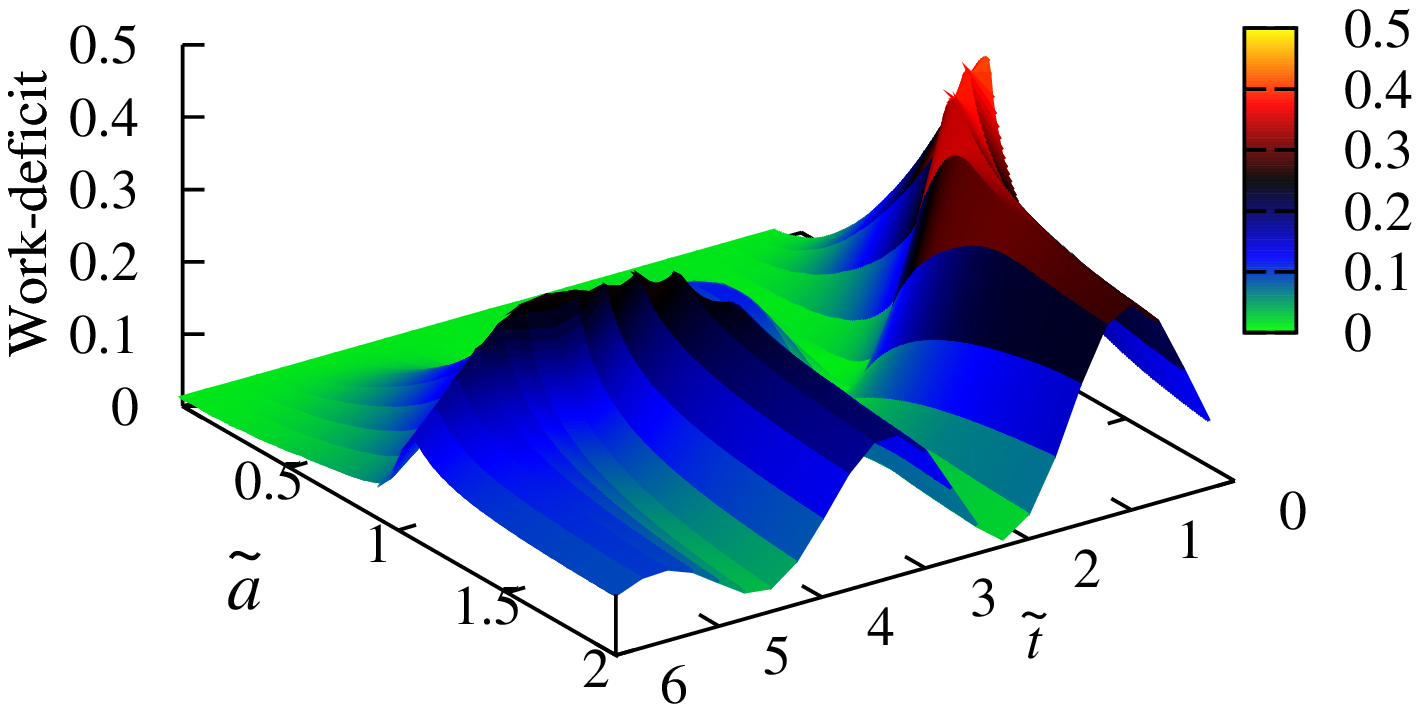}}\hspace{0.3cm}
\caption{(Color online) Behavior of entanglement  and  quantum work-deficit with respect to time and transverse field. Entanglement, as quantified by logarithmic negativity, is plotted in panels (a) and (c), and 
quantum work deficit is plotted in panels (b) and (d), 
as functions of the evolution time $\tilde{t}$ and initial field strength $\tilde{a}$, with $\gamma = 0.4$ (top row) and $\gamma = 0.6$ (bottom row). 
Entanglement and quantum work-deficit are respectively measured in ebits and qubits. The other axes are dimensionless. 
The revival of entanglement is seen to be directly related to the amount of work-deficit present at any particular time for a fixed $\gamma$, viz. the times 
for which the quantum work-deficit is significantly high as a function of $\tilde{a}$ correspond to the times at which entanglement revival has taken place. }
\label{fig7}
\end{center}
\end{figure*}

\section{Measures of quantum correlation}
\label{s2}

\noindent\emph{$\textit{Entanglement (Logarithmic Negativity)}$}: For analyzing the properties of bipartite entanglement of the nonequilibrium time-evolved
state of the anisotropic XY spin chain, the logarithmic negativity (LN) proves to be a useful computational measure of entanglement \cite{12}. The definition of logarithmic negativity is based on the Peres-Horodecki separability criterion \cite{13, 14}. The negativity of the partial transpose of any two-party state is a sufficient condition for bipartite entanglement. The condition is necessary and sufficient for two qubits \cite{14}.
%
%
The LN of an arbitrary bipartite state, \(\rho_{12}\), is defined as
\begin{eqnarray}
E_{\cal{N}}(\rho_{12})= \log_2 \left\|\rho_{12}^{T_A}\right\|_1
\equiv  \log_2 [ 2 \cal{N}(\rho_{12}) +1 ],
\end{eqnarray}
where
$\cal{N}(\rho_{12}) = (1/2)(\left\|\rho_{12}^{T_1}\right\|_1- 1)$
is called the ``negativity", and $\left\|\rho_{12}^{T_1}\right\|_1$ is the trace norm of the partially  transposed state, \(\rho_{12}^{T_1}\), of \(\rho_{12}\). 
The negativity, $\cal{N}(\rho_{12})$, is thus the sum of the absolute values of the negative eigenvalues of \(\rho_{12}^{T_1}\). 
For general mixed states, the LN is an upper bound for distillable entanglement \cite{12}.\\

\noindent\emph{$\textit{Quantum work-deficit}$}: Quantum work deficit (QWD) is based on the concept that information is a thermodynamic resource \cite{10} and its utility and dynamics is governed by similar laws.
It is defined as the difference between the amount of extractable pure states under suitably restricted global and local operations \cite{QWD}. 
Hence, QWD is an information-theoretic measure of quantum correlation, independent of the entanglement-separability criteria.

The allowed class of global operations on a quantum state $\rho_{12}$ is called ``closed operations" (CO). This set of operations
consists of: (a) unitary operations, and (b) dephasing the quantum state $\rho_{12}$ using an orthonormal projector set $\{X_i\}$ such that $\rho_{12}\rightarrow \sum_i X_i \rho_{12} X_i$, where the operations are defined on the Hilbert space ($\cal{H}$) of $\rho_{12}$. The amount of pure states extractable under CO can be shown to be $I_{CO}=N-S(\rho_{12})$. Here, $N$ = $\log_2$ dim $(\cal{H})$ and $S(\rho)$ denotes the von Neumann entropy of the state $\rho$. For the case of local operations, the allowed class of operations is closed local operations and classical communication (CLOCC) which consists of: a) local unitary operations b) dephasing locally and communicating a dephased subsystem to the other party over a noiseless quantum channel. The amount of pure states extractable under CLOCC is given by $I_\textrm{CLOCC}=N- \inf_{\Lambda \in \textrm{CLOCC}}[S(\rho'_{1})+ S(\rho'_{2})]$, where 
$
\rho'_i= \Tr_j(\Lambda(\rho_{ij})) (i,j=1,2; i \neq j)
$.
The work deficit ($\Delta(\rho_{12})$) is then given by
\begin{equation}
 \Delta(\rho_{12}) = I_\textrm{CO}(\rho_{12}) - I_\textrm{CLOCC}(\rho_{12}).
\end{equation}

The QWD can also be defined with respect to two-way communication of the dephased subsystem, although, it is hard to compute for arbitrary quantum states. Hence, we restrict ourselves to the case where one-way communication is permitted. For two-qubit systems, the dephasing involves a projection-valued measurement by projecting onto a single-qubit orthonormal basis. The most general form of the basis contains the states
$
\left|i_1\right\rangle= \cos \frac{\theta}{2} \left|0\right\rangle + e^{i\phi} \sin \frac{\theta}{2} \left|1\right\rangle$ and
$\left|i_2\right\rangle =  -e^{-i\phi} \sin \frac{\theta}{2} \left|0\right\rangle + \cos \frac{\theta}{2} \left|1\right\rangle$,
where \(\{|0\rangle, |1\rangle\}\) form the computational qubit basis.
%
%

\begin{figure}
\begin{center}
\label{fig1}
\subfigure[]{\includegraphics[width=.245\linewidth,clip=]{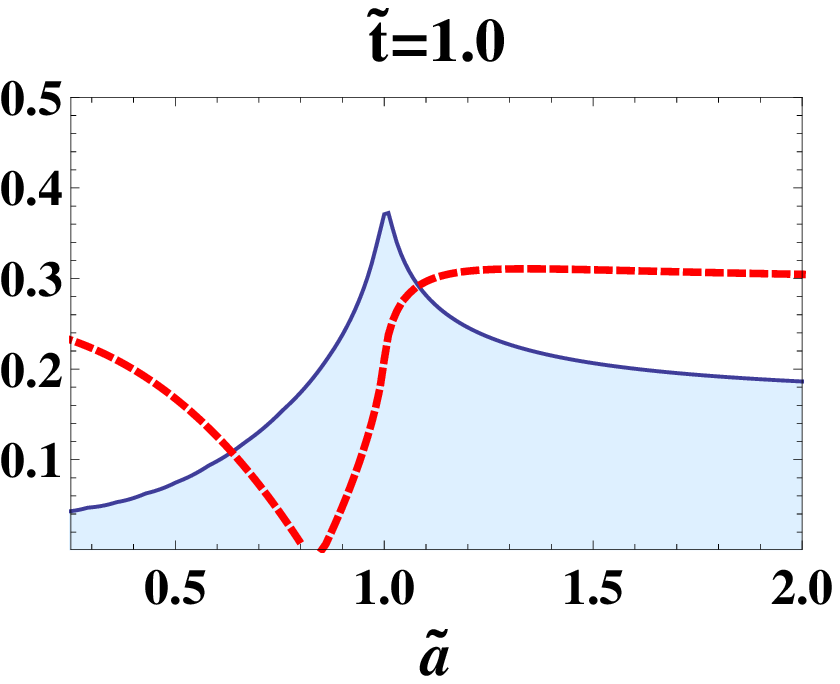}}\hspace{0.2cm}
\subfigure[]{\includegraphics[width=.245\linewidth,clip=]{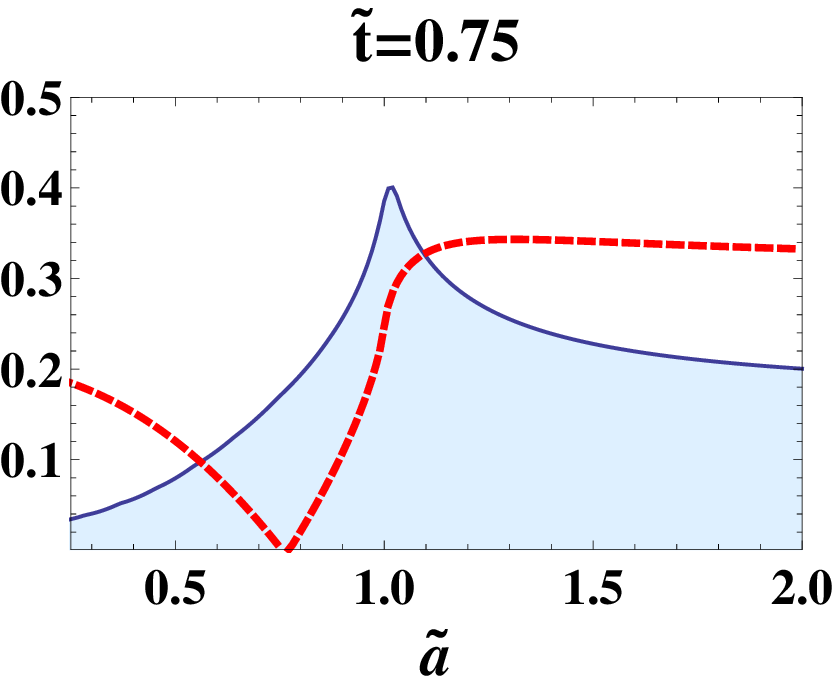}}\hspace{0.2cm}
\subfigure[]{\includegraphics[width=.245\linewidth,clip=]{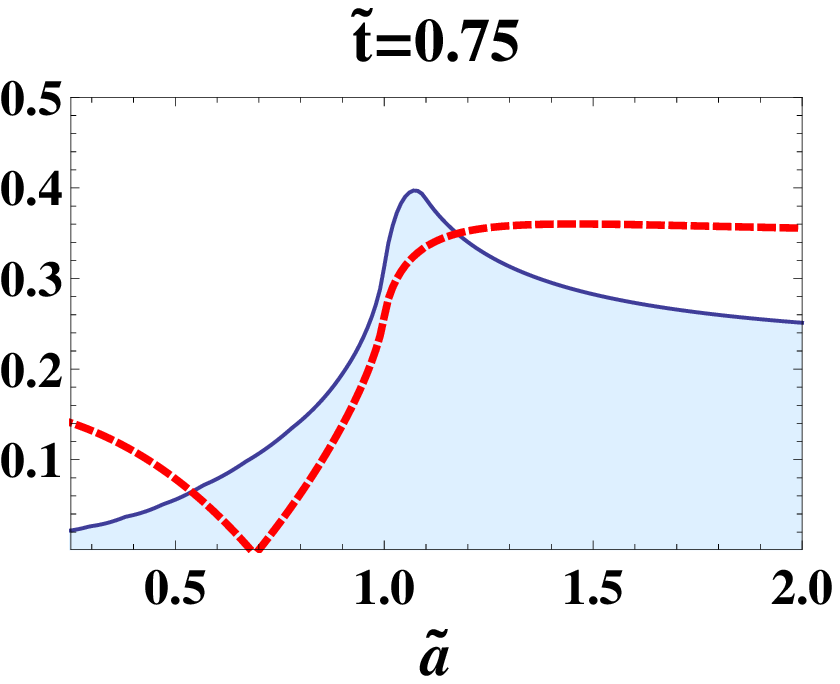}}\hspace{0.2cm}
\subfigure[]{\includegraphics[width=.245\linewidth,clip=]{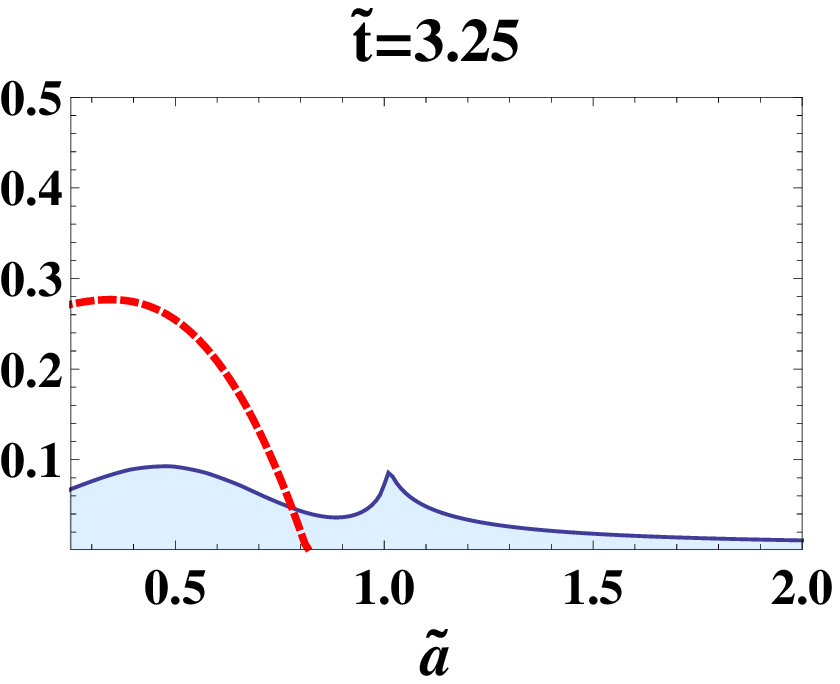}}\hspace{0.2cm}
\subfigure[]{\includegraphics[width=.245\linewidth,clip=]{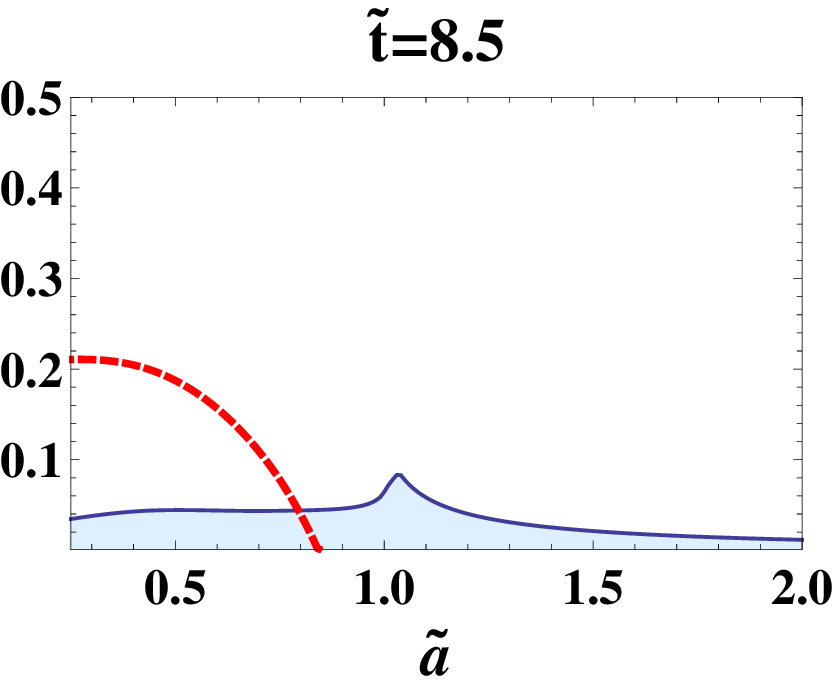}}\hspace{0.2cm}
\subfigure[]{\includegraphics[width=.245\linewidth,clip=]{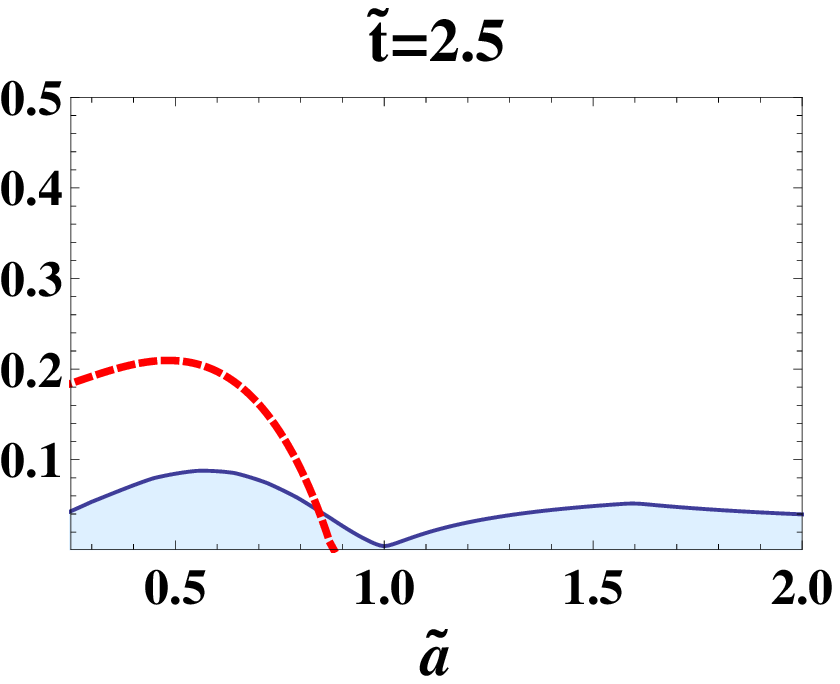}}\hspace{0.2cm}
\caption{(Color online) Behavior of entanglement (dashed line), measured in ebits, and quantum work-deficit (continuous line), measured in qubits, at fixed times $\tilde{t}$ as functions of the initial field strength $\tilde{a}$, for $\gamma = 0.4$ ((a) and (d)), $\gamma = 0.5$ ((b) and (e)), and $\gamma = 0.6$ ((c) and (f)). $\tilde{a}$ and $\tilde{t}$ are dimensionless parameters. It is seen that entanglement resurrection, after death, takes place only when the area (light blue) under the quantum work-deficit is significantly large. Resurrection takes place in the top panels ((a), (b) and (c)) as opposed to the bottom panels ((d), (e) and (f)) for the same $\gamma$ in which no revival occurs.}
\label{fig7-noi-eita}
\end{center}
\end{figure}

\section{Dynamical behavior of the correlation measures}
\label{s3}

\noindent\emph{$\textit{Entanglement Revival and Quantum Work-Deficit}$}: The time-evolved, nearest-neighbor bipartite entanglement, as measured using logarithmic negativity, is known to undergo a dynamical phase transition \cite{11} when considered with respect to the initial transverse field parameter $\tilde{a}$, for fixed evolution time $\tilde{t}$ and anisotropy parameter $\gamma$. The initial equilibrium state of the system is considered to be at $T$ = 0. The critical behavior of NN entanglement is observed for different values of the anisotropy parameter $\gamma$. The DPT of bipartite entanglement is also observed for other computable measures of entanglement such as concurrence \cite{con}. For two-qubit systems that we are considering here, concurrence and LN are zero if and only if the state is separable \cite{13, 14, con, add}.
Interestingly, however, we find that there is no DPT when the time-dynamics of the NN quantum work deficit is considered for the same system parameters.

In Fig.\,1, the dynamics of the two measures of quantum correlations with respect to evolution time ($\tilde{t}$) and the initial field parameter ($\tilde{a}$) is depicted. 
The figures on the left (Fig.\,1(a), (c)) give the behavior of LN, while on the right (Fig.\,1(b), (d)), we have the behavior of QWD, for anisotropy parameter values $\gamma$=0.4 (top) and $\gamma$=0.6 (bottom).
The DPT of entanglement, leading to entanglement death and possible revival or its absence, with changing initial field parameter, is closely associated with the dynamics of QWD. Though no DPT is observed for QWD, it is clear from Fig.\,1 
that the behavior of QWD is qualitatively different
at times when entanglement revival with respect to $\tilde{a}$ is observed, from the times when such revival is absent.

Figure 2 shows the behavior of LN and QWD of the time-evolved state, at different fixed times ($\tilde{t}$), as a function of the initial field parameter $\tilde{a}$.
%
The six panels in the figure pertain to different fixed times and each row of the figure corresponds to a particular value of the anisotropy parameter ($\gamma$ = 0.4, 0.5, 0.6). An important observation from the figure is that for certain values of evolution time (Fig.\,2 (a), (c) and (e)), entanglement death occurs at a certain initial field value, but entanglement again revives to give nonzero values at some higher initial field.
However, for the same anisotropy parameter but different fixed values of evolution time (Fig.\,2 (b), (d) and (f)), entanglement death occurs but does \emph{not} revive with the increase of initial magnetic field.
In particular, we observe that at times where revival of entanglement have taken place, the values of QWD are much higher than the times where revival does not occur. 
Hence, there is an indication that the dynamics of entanglement can be quantified by the behavior of QWD even in the absence of a direct operational relation.
For cases where entanglement revival occurs, the area under the QWD is much larger than for cases where revival is absent. 
We find that this feature is generic and is observed for the whole range of the anisotropy parameter, $\gamma \neq$ 0. 
Figures 1 and 2 qualitatively indicate that the behavior of entanglement and QWD is non-trivially related for the model under consideration. Below, we analyze the quantitative aspect of this feature. \\

\noindent\emph{$\textit{Quantitative Relation between Entanglement and QWD}$}: The property of QWD that fosters the dynamical behavior of entanglement can be estimated from Figs.\,1 and 2. We infer that the revival of entanglement is directly related to the area of the region under QWD as a function of the initial field strength $\tilde{a}$, for a fixed time $\tilde{t}$ and anisotropy $\gamma$. To obtain an analytical expression for this relation, we calculate the area under the QWD in the panels of Fig.\,2 under a suitable scaling.
By studying the dynamics of QWD and LN for different anisotropy values $\gamma$ and evolution times $\tilde{t}$, we find that on varying the initial field parameter $\tilde{a}$, the quantum correlation measures, QWD and LN, converge to finite values for $\tilde{a}\geq$ 2.0. The DPT is observed in the vicinity of $\tilde{a} \approx$ 1.0 and the most interesting dynamics of the quantum correlations occurs around this region. Let $\textit{A}^{\gamma}(\tilde{t})$ be the area under the plot of QWD for any particular time $\tilde{t}$ and anisotropy $\gamma$, for the initial field parameter ranging from $\tilde{a}$ = 0.0 to $\tilde{a}$ = 2.0.

We find that \emph{given a known area $\textit{A}^{\gamma}(\tilde{t})$ for which a revival of entanglement occurs, any situation with a higher area, $\textit{A}^{\gamma}(\tilde{t'})$, will always result in revival of entanglement.}
More specifically,
\begin{equation}
\textit{A}^{\gamma}(\tilde{t'}) \ge \textit{A}^{\gamma}(\tilde{t}) \Longrightarrow \textrm{Entanglement revival for $\tilde{t'}$},
\label{ss}
\end{equation}
provided $\textit{A}^{\gamma}(\tilde{t})$ is known to provide a revival. Hence, one is able to obtain a quantitative expression in terms of
QWD that can predict the revival of entanglement in an infinite spin system for a fixed $\gamma$, but at different $\tilde{t}$ provided a known case of revival is already obtained for a single set of parameters.
Furthermore, we now define a scale factor $\textit{S}^{\gamma}(\tilde{t})$ as
\begin{equation}
S^{\gamma}(\tilde{t}) \equiv \frac{A^{\gamma}(\tilde{t}) - A^{\gamma}_{min}}{M},
\label{s}
\end{equation}
where $\textit{A}^{\gamma}_{min}$ is the lowest value of $\textit{A}^{\gamma}(\tilde{t})$ for which a revival is already known. $\textit{M}$ is an arbitrary scaling parameter, which e.g. can be chosen to be \(A^\gamma_{min}\).
We call $S^{\gamma}(\tilde{t})$ the cumulative QWD, noting its similarity to cumulative frequency in a probability distribution.
The values of $S^{\gamma}(\tilde{t})$ for different times are plotted in Fig.\,3. It is evident from the panels in Fig.\,3 (for different $\gamma$) that the following holds:
\begin{eqnarray}
S^{\gamma}(\tilde{t}) &\geq& 0 \Longrightarrow \mbox{revival of entanglement occurs}, \nonumber\\
S^{\gamma}(\tilde{t}) &<& 0 \Longrightarrow \mbox{no revival}.
\label{t}
\end{eqnarray}
The estimation of $\textit{A}^{\gamma}_{min}$ requires an optimization. However, relations (\ref{s}) and (\ref{t}) can be used for a non-optimized $\textit{A}^{\gamma}_{min}$, which in fact leads to relation (\ref{ss}).
Comparing Figs.\,1 and 2 with Fig.\,3, we find that \(S^\gamma(\tilde{t})\), which is a function of the QWD, correctly signals the time of entanglement death and resurrection for different \(\gamma\). 
The behavior of \(S^\gamma(\tilde{t})\) also indicates that with the increase of \(\gamma\), entanglement revival becomes more and more frequent. In particular, when \(\gamma\) is close to unity so that the system  is close 
to the transverse Ising model, the characteristics of \(S^\gamma(\tilde{t})\) point to an almost periodic entanglement death and resurrection. 



%
%
\begin{figure}
\begin{center}
\label{fig.3}
\subfigure[]{\includegraphics[width=.245\linewidth,clip=]{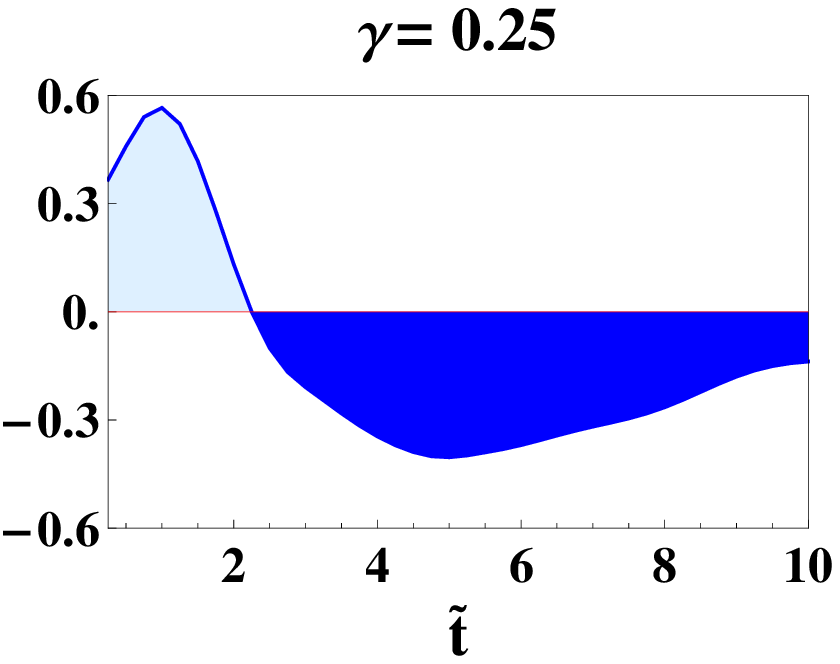}}\hspace{0.2cm}
\subfigure[]{\includegraphics[width=.245\linewidth,clip=]{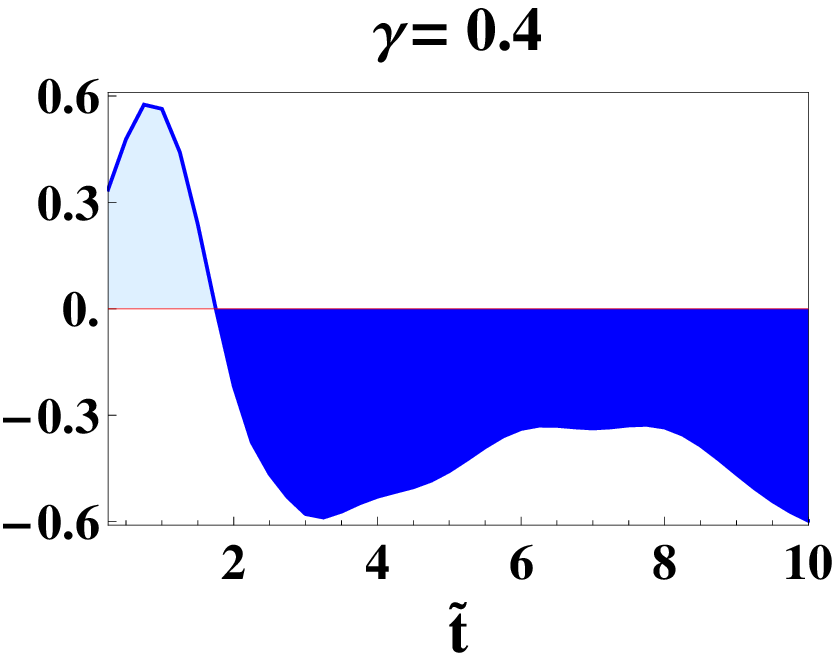}}\hspace{0.2cm}
\subfigure[]{\includegraphics[width=.245\linewidth,clip=]{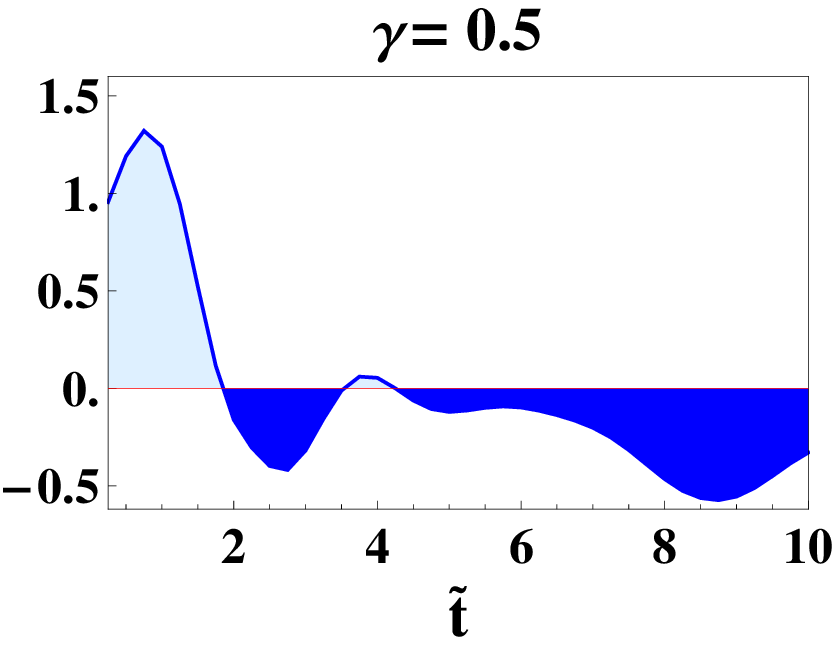}}\hspace{0.2cm}
\subfigure[]{\includegraphics[width=.245\linewidth,clip=]{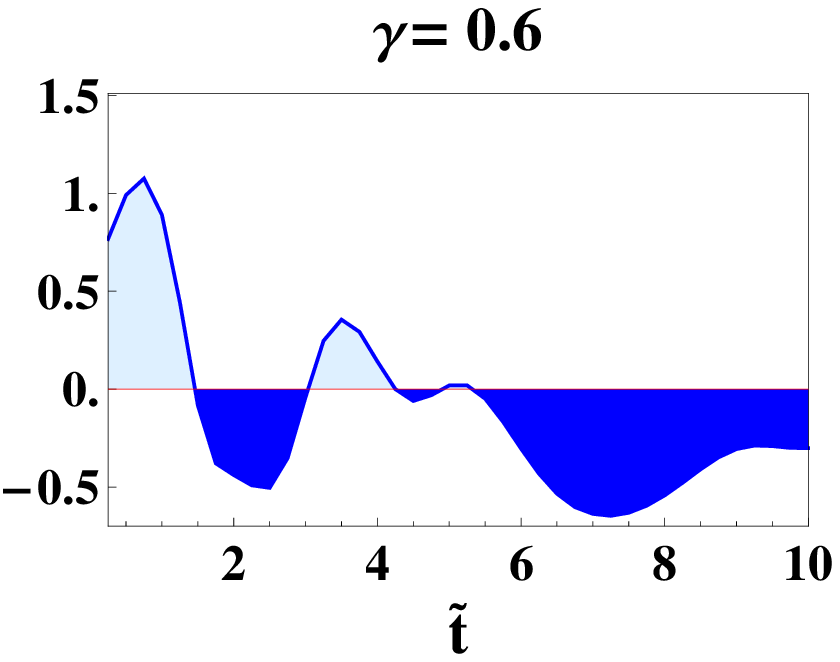}}\hspace{0.2cm}
\subfigure[]{\includegraphics[width=.245\linewidth,clip=]{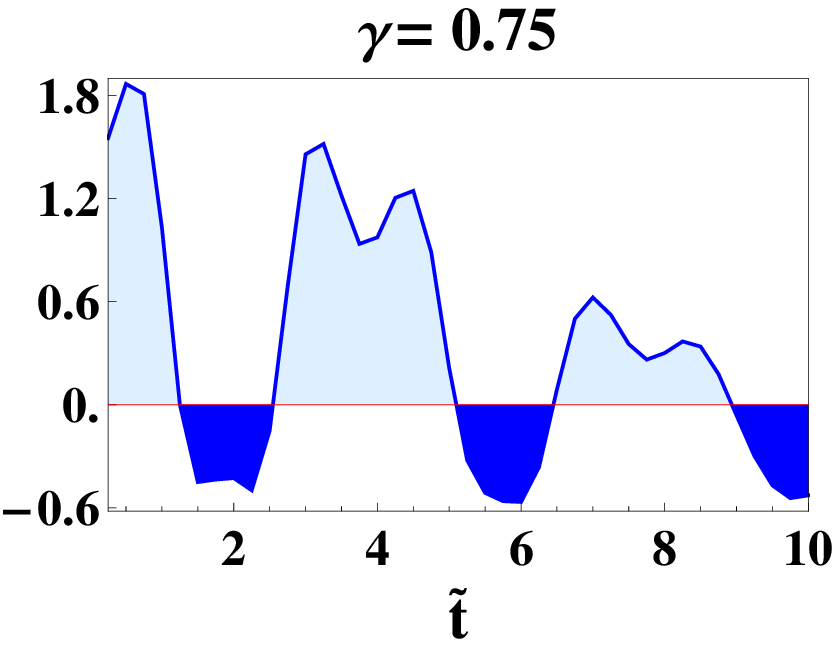}}\hspace{0.2cm}
\subfigure[]{\includegraphics[width=.245\linewidth,clip=]{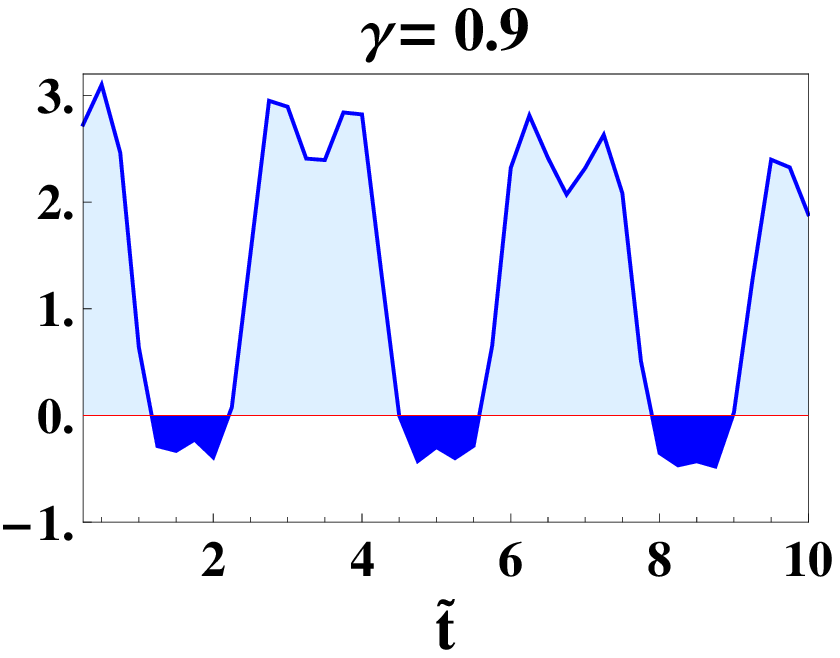}}\hspace{0.2cm}
\caption{(Color online) Behavior of the scale factor $\textit{S}^{\gamma}(\tilde{t})$ as a function of time $\tilde{t}$ for different values of the anisotropy parameter $\gamma$. For a given $\gamma$, the non-negative part (the region of $\tilde{t}$ for which the scale factor is non-negative) corresponds to the region (light blue) where entanglement revival is observed, while the negative region (dark blue) corresponds to no-revival. For the above plots, the scaling parameter $M$ is equal to $A^\gamma_{min}$. $\textit{S}^{\gamma}(\tilde{t})$, $\tilde{t}$ and $\gamma$ are all dimensionless.}
\end{center}
\end{figure}
%
%

An interesting feature of the dynamics is the sudden quantum quenching of the transverse field $h(t)$. The initial field, $h(0) = \tilde{a}$, 
is quantum quenched to $h(t) = 0$ for $t >$ 0. For certain values of $\tilde{a}$, the sudden quench could involve a rapid passage of the non-equilibrium system across a quantum critical point \cite{sachnew}. 
It may be noted here that the critical values of $\tilde{a}$ and the anisotropy parameter $\tilde{\gamma}$ could also be determined by the differential convertible phases of the anisotropic XY model \cite{DLC}. 
For rapid quenches that cross the critical field $\tilde{a}_c$, the dynamics of quantum correlations are significantly different from those quenches that 
do not cross any critical point. This can be seen from Fig.\,1 for anisotropy values $\tilde{\gamma} = 0.4$ and $\tilde{\gamma} = 0.6$. For fixed values of $\tilde{a}$, the temporal dynamics of the correlations is different 
above and below the critical point $\tilde{a}_c$.
Such qualitatively different behavior as the system passes through criticality can be used to study the critical scaling of the quantum correlations 
under finite quenching and the Kibble-Zurek mechanism \cite{KZ}.
This is also potentially important to results pertaining to quenched quantum dynamical phase transition \cite{polka} and critical scaling \cite{huse}.

\section{Discussion}
\label{s4}

Quantum correlations of shared systems can broadly be divided into two categories -- ones that are conceptualized by using the
entanglement-separability criteria, and the ones that are done using information-theoretic concepts. For general mixed states, there exist no direct operational relation between these categories of correlation measures. In this work, we attempt to create a quantitative inter-relation between two specific measures from each of the mentioned categories, namely, 
the logarithmic negativity, an entanglement measure, and the quantum work-deficit, an information-theoretic quantum correlation measure. We obtain a quantitative relation between these two measures in the time-dynamics of the infinite anisotropic quantum XY spin-1/2 chains in an external perturbative transverse magnetic field. Entanglement is known to exhibit a dynamical phase transition in the time-evolution of the system: for certain evolution times, the entanglement vanishes for a certain value of the initial magnetic field and gets resurrected again, while for certain other times, the entanglement vanishes and does not become non-negative later. We find that the resurrection in entanglement, for a given evolution time, is related to the area accumulated under the quantum work-deficit curve as a function of the initial magnetic field, for that time. An accumulated area under the 
quantum work-deficit curve that is higher than a certain threshold, ``forces'' the resurrection of entanglement to occur.
%
The relation can be shown in the form of a quantitative relation that captures the dependence of the entanglement behavior on the dynamics of QWD.
The quantitative relation that is obtained between the two quantum correlation measures for the anisotropic quantum XY spin-1/2 models can
%
%
possibly be extended to other generic quantum many-body systems to obtain a better understanding of nonclassicality in those models. 
The results may also prove useful in developing a robust operational relation between the two categories of correlation measures.
Further, the quantum quenching of the transverse field can be used to study the scaling of quantum correlations across critical points in dynamical phase transitions.

\section*{Acknowledgments}

The work of HSD is supported by the University Grants Commission (UGC), India. HSD thanks the Harish-Chandra Research Institute (HRI) for hospitality and support during visits. 
We acknowledge computations performed at the cluster computing facility in HRI and also at the UGC-DSA computing facility at the Jawaharlal Nehru University.

\end{document}